\documentclass[nofootinbib,floatfix,preprintnumbers,
amsmath,amssymb,
onecolumn]{revtex4}
\usepackage{epsfig}

\newcommand{\GeV}{\,\mathrm{GeV}}

\newcommand{\beq}{\begin{equation}}
\newcommand{\eeq}{\end{equation}}
\newcommand{\bea}{\begin{eqnarray}}
\newcommand{\eea}{\end{eqnarray}}
\newcommand{\beqn}{\begin{eqnarray}}
\newcommand{\eeqn}{\end{eqnarray}}
\newcommand{\beas}{\begin{eqnarray*}}
\newcommand{\eeas}{\end{eqnarray*}}

\newcommand{\bquo}{\begin{quote}}
\newcommand{\enqu}{\end{quote}}

\def\be{\begin{equation}}
\def\ee{\end{equation}}
\def\bea{\begin{eqnarray}}
\def\eea{\end{eqnarray}}
\begin{document}
\title{ Superpartner spectrum of minimal gaugino-gauge mediation}
\author{Roberto Auzzi and Amit Giveon}
\email{auzzi@phys.huji.ac.il, giveon@phys.huji.ac.il}
\affiliation{Racah Institute of Physics, The Hebrew University, Jerusalem 91904, Israel}
\begin{abstract}
We evaluate the sparticle mass spectrum in the minimal four-dimensional construction
that interpolates between gaugino and ordinary gauge mediation at the weak scale.
We find that even in the hybrid case -- when the messenger scale is comparable to
the mass of the additional gauge particles -- both the right-handed as well as the
left-handed sleptons are lighter than the bino in the low-scale mediation regime.
This implies a chain of lepton production and, consequently,
striking signatures that may be probed at the LHC already in the near future.

\end{abstract}
\maketitle


\section{Introduction}

In this note, we compute the sparticle spectrum in the minimal
construction   \cite{Cheng,Csaki} of ``gaugino mediation''
\cite{KKS,Chacko}   at the weak scale.
Our renewed interest in these models
(which are a particular example of General Gauge Mediation \cite{GGM}) is twofold.
First, this class of models is expected to have interesting phenomenological
properties, which may allow early discovery at LHC,
in some regime of its parameters space \cite{dsfs1,dsfs2}.
Second, simple generalizations of such models have a natural embedding in (deformed) SQCD,
and can thus provide a dynamical realization of direct gaugino mediation \cite{gkk}.

This minimal four-dimensional construction interpolates between gaugino
and ordinary gauge mediation, and we thus refer to it as
``Minimal gaugino-Gauge Mediation'' (MgGM).
In ref. \cite{mggmsoft}, we computed the sparticle mass spectrum of MgGM at the messenger scale
(see also  \cite{McGarrie,Sudano} for generalizations).
The theoretical setting and the explicit results of
\cite{mggmsoft} will be presented in section 2.
In section 3 of the present work, we evaluate the sparticle spectrum
in this class of models at the weak scale.

The main result of this note is the following.
Even in the hybrid case -- when the messenger scale is comparable to
the mass of the additional gauge particles -- both the right-handed as well as the
left-handed sleptons are lighter than the bino in the low-scale mediation regime.
This implies a chain of lepton production and, consequently, striking signatures
that may be probed at the LHC already in the near future   \cite{dsfs1,dsfs2}.
Our results are further discussed in section 4.

\section{Theoretical setting}

\begin{figure}[h]
 \centerline{\includegraphics[width=3in]{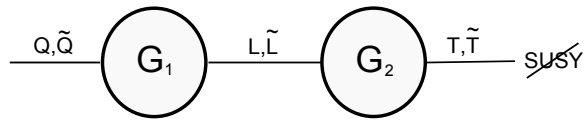}}
 \caption{\footnotesize Quiver diagram for the model. }
\label{quiver}
\end{figure}

The model consists of a visible sector
with  the MSSM matter fields $(Q,\tilde{Q})$, which are charged under a gauge group $G_1$;
the messenger fields $(T,\tilde{T})$ are charged instead under a second gauge group $G_2$.
Supersymmetry breaking is communicated to the visible sector
 by link fields $(L,\tilde{L})$, which are charged under both of the gauge groups.
A quiver diagram for the model is shown in figure \ref{quiver}.
The messenger fields $(T,\tilde{T})$ couple to the spurion of SUSY-breaking, $S$,
whose scalar components get VEVs,
\beq
S=M+\theta^2 F \, , \label{sss}
\eeq
as in Minimal Gauge Mediation (MGM),
and we denote the effective SUSY-breaking scale, $F/M$,  by
\beq
\Lambda\equiv F/M\, . \label{lll}
\eeq
Finally, the link fields get a VEV,
\beq
L=\tilde L=v\, .
\eeq
This model interpolates between gaugino mediation
(when the VEV is much smaller than the messenger scale, $v\ll M$),
and MGM (when $v\gg M$).

The gauge groups $G_1$ and $G_2$ are both chosen to be $SU(5)$;
at energies below the gauge coupling unification scale,
$G_{1}$ is spontaneously broken to $SU(3) \times SU(2) \times U(1)$.
The VEV of the link fields $v$ breaks  $G_{1} \times G_{2}$ to
a linear combination of the two groups, with gauge couplings
 \beq
\frac{1}{\left(g_{SM}^{(r)}\right)^2}=
\frac{1}{\left(g_1^{(r)}\right)^2 }+\frac{1}{\left(g_2^{(r)}\right)^2 } \, , \label{gsm}
\eeq
where $r=1,2,3$ correspond to the $U(1)$, $SU(2)$, $SU(3)$
gauge couplings, respectively.

Let us introduce the four dimensionless parameters $x$ and $y_r$:
\beq
x \equiv{\Lambda\over M}~,\qquad
y_r \equiv{m_{v_r}\over M}~,  \qquad
 m_{v_r}= 2v\sqrt{\left(g_1^{(r)}\right)^2+\left(g_2^{(r)}\right)^2} \, ,
  \label{xxyy}
\eeq
where $m_{v_r}$ is the mass of the corresponding massive combination of the
gauge particles of $G_{1} \times G_{2}$,
which is spontaneously broken by $v$.
In order to maintain the gauge coupling unification
we require that $G_2$ is $SU(5)$-invariant
 just above the scale $v$ \cite{Cheng};
  then all the three couplings $g_2^{(r)}$
are equal to the same value $g_2$.
In this work we consider cases where $M>v$.
The mass scales,
\beq
m_{v_r}=\frac{2 v g_2^2}{ \sqrt{  g_2^2 - (g_{\rm SM}^{(r)})^2}} \, ,
\eeq
at the messenger scale $M$, are thus functions of
the coupling constant $g_2$ and the MSSM gauge couplings,
measured at the messenger scale.

\begin{figure}[h]
 \centerline{\includegraphics[width=3in]{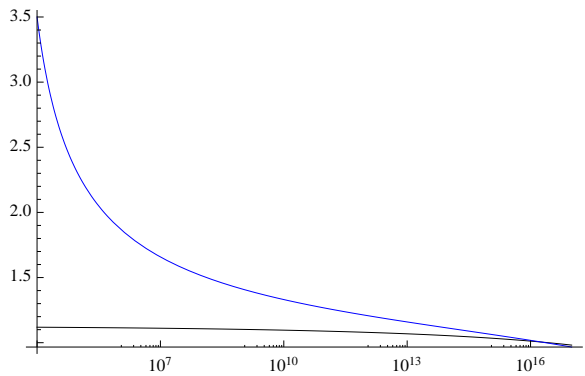}}
 \caption{\footnotesize Plot of $m_{v_2}/m_{v_1}$ (lower curve) and $m_{v_3}/m_{v_1}$ (upper curve)
 as a function of the messenger scale, for $\alpha_{g_2}^{-1}(M)=10$.  }
\label{ratiomv}
\end{figure}

The coupling $g_2$ is a free parameter, which controls how much
the sfermions mass suppression factor is uniform among the three SM gauge groups;
in the formal limit $g_2\gg g_{\rm SM}^{(3)}$, the suppression factors
are the same for each gauge group. In the opposite limit,
where  $g_2 \approx  g_{\rm SM}^{(3)}$,
there is almost no suppression factor for the squarks masses,
while there is  suppression for the sleptons masses.
For the model to be calculable, we cannot choose $g_2$ to be too strong.
The quantities $(m_{v_2}/m_{v_1},m_{v_3}/m_{v_1})$ as
 a function of the energy scale are plotted in figure \ref{ratiomv},
 for $\alpha_{g_2}^{-1}(M)=10$.

The gauginos soft masses are the same as in minimal gauge mediation \cite{Martin1996,Dimopoulos1996}:
\beq
m_{\tilde{g}_r}=\frac{\alpha_{SM}^{(r)}}{4 \pi} \Lambda n_r \, q(x) \, ,
\qquad  \alpha_{SM}^{(r)}\equiv\frac{\left(g_{SM}^{(r)}\right)^2 }{4 \pi } \, ,
\label{mgauginos}
\eeq
where
\beq
q(x)=\frac{1}{x^2} \left( (1+x ) \log(1+x) + (1-x) \log(1-x) \right) \, ,  \label{gauginomass}
\eeq
and $n_r$ is the Dynkin index for the pair of messengers
in a normalization where $n_r=1$  for $N+\bar{N}$ of $SU(N)$, and $n_1={6\over 5}Y^2$
for a messenger pair with electro-weak hypercharge $Y=Q_{\rm EM}-T_3$
(we use the GUT normalization for $\alpha_1$, as in \cite{Martin1996}).

The sfermions soft masses  were computed in \cite{mggmsoft}:
\beq
m_{\tilde{f}}^2=2 \Lambda^2 \sum_r  \left( \frac{     \alpha_{SM}^{(r)}    }{4 \pi} \right)^2
C_r^{\tilde{f}}n_r s(x,y_r) \, , \label{mssmres}
\eeq
where $C_r^{\tilde{f}}$ is the quadratic Casimir invariant of
the MSSM scalar field $\tilde{f}$, in a normalization where
$C_3=4/3$ for color triplets, $C_2=3/4$ for $SU(2)$ doublets and
$C_1={3\over 5}Y^2$. The function $s(x,y)$ is given by
\beq
s(x,y)={1\over 2x^2}\left(s_0 +\frac{s_1+ s_2}{y^2} + s_3 +s_4 +s_5 \right)
 + \, (x \rightarrow -x)\, , \label{esse}
\eeq
where
\beq
s_0=2(1+x) \left( \log (1+x) -2 {\rm Li}_2  \left( \frac{ x}{1+x}\right)
+\frac{1}{2} {\rm Li}_2 \left( \frac{2 x}{1+x}\right) \right)   \, , \label{mgmmasses2}
\eeq
\[
s_1=- 4 x^2  - 2 x(1+x) \log^2(1+x) - x^2 \, {\rm Li}_2(x^2) \, ,
\]
\[
s_2=8 \left(1+x\right)^2 h\left(\frac{y^2}{1+x},1\right)-4 x \left(1+x\right) h\left(\frac{y^2}{1+x},\frac{1}{1+x}\right)
   -4 x    h\left(y^2,1+x \right)-8 h\left(y^2,1\right) \, ,
\]
\[
s_3= -2 h\left(\frac{1}{y^2},\frac{1}{y^2}\right)
-2 x \,   h\left(\frac{1+x}{y^2},\frac{1}{y^2}\right) +
2(1+ x) h\left(\frac{1+x}{y^2},\frac{1+x}{y^2}\right) \, ,
\]
\[
s_4=(1+x) \left(  2  h\left(\frac{y^2}{1+x},\frac{1}{1+x}\right)
 - h\left(\frac{y^2}{1+x},1\right)- h\left(\frac{y^2}{1+x},\frac{1-x}{1+x}\right)  \right) \, ,
\qquad
s_5= 2 h\left(y^2,1+x\right)-2 h\left(y^2,1\right) \, .
\]
The function $h$ is defined by the integral  \cite{veltman}:
\beq
h(a,b)= \int_0^1 dx  \left( 1+ {\rm Li}_2 (1-\mu^2) -\frac{\mu^2}{1-\mu^2}  \log \mu^2 \right) \, ,
\qquad \mu^2=\frac{a x + b(1-x)}{x(1-x)} \, ,\label{hab}
\eeq
where the dilogarithm is defined by ${\rm Li}_2(x)=-\int_0^1 {dt\over t}\log(1-xt)$,
and an analytical expression for $h$ can be found in \cite{veltman}.
We refer the reader to ref. \cite{mggmsoft} for more details.

\begin{figure}[h]
 \centerline{\includegraphics[width=3in]{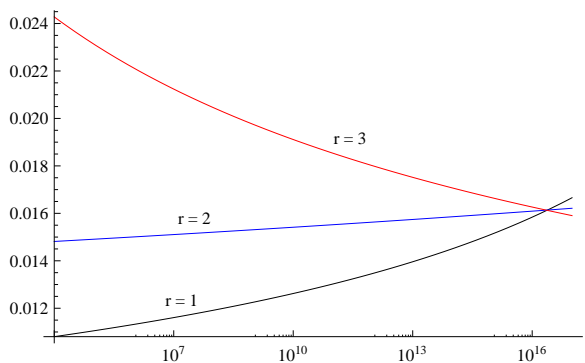}}
 \caption{\footnotesize The quantity  $\omega_r=\alpha_{SM}^{(r)}/(4 \pi)$
 plotted as a function of energy; this gives an estimation of the three loops corrections. }
\label{omega}
\end{figure}

Finally, a comment is in order. Equations~(\ref{mssmres})--(\ref{hab})
give the sfermions masses in the two-loop approximation.
As pointed out in \cite{gkk}, in some regimes of parameters space,
where the suppression factors $s(r)\equiv s(x,y_r)$ are small, it might be that
the three-loop corrections are actually bigger than the two-loop ones.
The size of the three-loop corrections is weighted by an
extra loop factor, $\omega_r\equiv\alpha_{SM}^{(r)}/(4 \pi)$,
instead of the suppression factors $s(r)$;
a plot of $\omega_r$ is given in figure \ref{omega}.
Hence, the higher-loops corrections are negligible
when $\omega_r$ are sufficiently smaller than $s(r)$;
for the examples presented in the following section,
we will argue that this is indeed the case.

\section{Sparticle spectrum }

If the massive gauge particles masses are bigger than the messenger scale,
$m_{v_r}\geq M$, for all $r=1,2,3$,
we can ignore contributions to the soft terms beta functions from the hidden sector.
In the following we will focus on some examples with $m_{v_1}=M$.
These are particularly interesting, since they provide a
hybrid model of MgGM -- something which is right in between gaugino mediation and MGM.
We consider for simplicity the case of  $N_{\rm mess}$ degenerate
messengers in the $(5+\bar{5})$ representation of $SU(5)$.

At one loop (and ignoring the threshold  corrections), the
MSSM gauge couplings as functions of the energy scale are:
\beq
\left(\alpha_{SM}^{(1)}\right)^{-1}=59.2-\frac{33}{5} \frac{ t}{2 \pi} \, , \qquad
\left(\alpha_{SM}^{(2)}\right)^{-1}=29.6-\frac{t}{2 \pi} \, , \qquad
\left(\alpha_{SM}^{(3)}\right)^{-1}=8.5+\frac{3 t}{2 \pi} \, ,  \label{asmasmasm}\eeq
where $t=\ln(Q/m_Z)$.
We use this as an initial estimation in order to compute
the soft masses at the messenger scale;
the program SOFTSUSY \cite{softsusy} is then used to
solve the Renormalization Group Evolution down to the electroweak scale
and to compute the physical spectrum.
These are used by SOFTSUSY to compute the
gauge couplings at the messenger scale; the result is then used
to correct the initial estimation for the soft masses.
This procedure is repeated until a self-consistent result is
found (typically convergence is achieved with good approximation after
$3-4$ steps).
The trilinear A-terms are set to zero at the messenger scale
(this is usually a good approximation because
they have the dimension of a mass and they are suppressed by 
an extra loop factor compared to the gaugino masses).
 We work in the usual approximation where only the
$(3,3)$ family components, $y_b,y_t,y_\tau$, of each
Yukawa couplings matrix are important
(for a review of these and other issues, see e.g. \cite{martin-review,mv,pbmz}).

Concerning the Higgs sector, as common in this kind of phenomenological studies \cite{mgma,mgmb},
we do not choose any specific model for $(\mu,B \mu)$,
but we treat them as free parameters.
The value of $\tan\beta$ is fixed at the beginning of the calculation;
the parameters $(\mu, B \mu)$ are then found from the Higgs VEV
and from the chosen value of $\tan\beta$.
We take the boundary conditions for the soft masses of
the Higgs fields, $m^2_{H_u},m^2_{H_d}$,
to be their values as obtained from eq.~(\ref{mssmres}),
namely, we assume that there are no extra contributions
from the mechanism that generates $(\mu, B \mu)$.

Some examples of the spectrum, for $\tan \beta=8, 20, 50$ and $N_{\rm mess}=1$, are shown in tables
\ref{masses-example-tan8}, \ref{masses-example-tan20}, \ref{masses-example-tan50}, respectively.
Some examples with $\tan \beta = 20$ and $N_{\rm mess}=5$ are shown
in table  \ref{masses-example-nmess5}.
In these tables we choose $\alpha_{g_2}^{-1} =10$; we have checked that
for $\alpha_{g_2}^{-1} =5$ there are no big differences in the spectrum.
For comparison, some examples of the spectrum in Minimal Gauge Mediation,
for $\tan \beta =20$ and $N_{\rm mess}=1,5$, are shown in tables \ref{masses-example-mgm},
 \ref{masses-example-mgm-nmess5}, respectively.

On the first column of each table the case of low messenger scale
$M \approx 10^5 {\rm GeV}$ is considered, with the requirement $M=0.99 \Lambda$;
this corresponds to a rather extreme corner of the parameters space,
where the lighter messenger has a mass near the 10 TeV,
and where the sfermion masses are particularly suppressed.
The precise value of $M$ is chosen in order to satisfy the
two experimental constraints:  $m_{h_0} > 114.4 \, {\rm GeV}$,
and that the mass of the lowest charged slepton
$m_{\tilde{\tau}_1}$ is bigger than about $100 \, {\rm GeV}$.
In this regime,
the suppression factors $s(1)$ and $s(2)$ are rather small,
comparable to the three-loop extra factor $\omega_r$ in figure 3.
We have checked though that allowing
for corrections in $s(r)$ of order $\omega_r$,
the results are essentially unchanged.
The $\tilde{\tau}_1$ slepton is mainly right-handed;
the lightest neutralino is mainly a bino in the case of one messenger,
while it is mainly higgsino in the case of $N_{\rm mess}=5$.

On the second and third columns, the cases of $M = 10^8,10^{15} \GeV$
are considered; the value of $\Lambda$
is chosen to satisfy the two constraints
on $m_{h_0}$ and $m_{\tilde{\tau}_1}$.
In all these cases, the lightest neutralino is mainly a bino,
and the $\tilde{\tau}_1$ slepton is mainly right-handed.
The three-loop corrections are negligible in these cases.

In the tables, in addition to the sparticle spectrum at the weak scale,
we also present the values of the soft masses
at the messenger scale, $M_r,m_Q,m_u,m_d,m_L,m_e,\mu,B\mu$,
as well as the messenger scale $M$ itself,
the effective SUSY-breaking scale $\Lambda$,
and the link fields VEV, $v$.
The values of the suppression factors $s(r)$,
and the parameters $y_r$, are also introduced in each example.
Note that the values of $y_r$ in the tables take into account 
the threshold corrections to $\alpha_{SM}(M)$ obtained in the iterations described above;
on the other hand, the plot in figure \ref{ratiomv} 
uses the input values in eq.~(\ref{asmasmasm}). 

Already for one messenger, in most of the parameters space
the NLSP is the $\tilde{\tau}_1$ slepton.
The exception to this is realized for a very large messenger scale $M$
and sufficiently low $\tan \beta$;
for $M=10^{15} \GeV$ and $\tan\beta=20$, the lightest neutralino
and the $\tilde{\tau}_1$ have a comparable mass.
This is different from Minimal Gauge Mediation,
where for one messenger the NLSP is mainly a bino in most of the parameters
space.

In the case of a low messenger scale $M\approx10^5 \GeV$, $\Lambda \geq 0.8 M$
with one messenger, both the right-handed and the
left-handed sleptons are lighter than the lightest neutralino.
If we increase the number of messengers, this kind of spectrum is much more generic;
for example, for $N_{\rm mess}=5$, this is generically possible for $M \leq 10^8 \GeV$
and $\Lambda \approx 10^5 \GeV$.
This scenario can lead to striking collider signatures
 \cite{dsfs1,dsfs2};
squark decay chains can pass through one or more sleptons and typical final states
from squark and gluino production at the LHC would include multiple leptons.

Finally, models with $B\mu=0$ at the messenger scale
may have phenomenologically attractive features,
e.g. because a vanishing $B\mu$ may provide an explanation
to the absence of potentially dangerous $CP$ violating phases $\mu^* (B \mu) M_r^*$,
and it may sometimes be useful in addressing the $\mu/B\mu$ problem.
It is thus curios to note that this can be achieved in MgGM,
for example for $N_{\rm mess}=1$, $\tan \beta=20$, $M \approx 2.615 \times 10^{13} \GeV$
and $\Lambda=1.7 \times 10^5$.
A  scan of the parameter space of
General Gauge Mediation models with $B\mu=0$ was recently
discussed in \cite{ADJK}.

\section{Discussion}

In this note, we computed the sparticle mass spectrum
in Minimal gaugino-Gauge Mediation (MgGM) at the weak scale
in various regimes in the parameters space (\ref{xxyy}),
and for several numbers of messengers.
In particular, we found that for a sufficiently low messenger scale $M$,
the NLSP is the right-handed stau, followed by the right-handed selectron,
and the left-handed sleptons, all of which are lighter than the lightest
neutralino.
Such a hierarchy of the spectrum has striking signatures in hadronic colliders,
since it gives rise to a cascade of leptonic production,
and consequently it may be
probed at the LHC in the near future \cite{dsfs1,dsfs2}.

It is remarkable that such a weak-scale phenomenology can be obtained already
in the minimal model, yet it should be interesting to investigate some
generalizations. For instance,
it would be instructive to inspect a general messenger sector,
along the lines of the General Messenger Gauge Mediation (GMGM) models
studied in~\cite{GMGM}
(note however that MgGM and its generalizations are not GMGM-type).
Of particular interest are the weak-scale phenomenological aspects
of the ``Direct Gaugino Mediation'' models of \cite{gkk}, and their
generalizations, since they have a simple dynamical realization in
deformed SQCD.

It should also be interesting to study the effects of
doublet-triplet splitting in the messenger sector,
as in \cite{Martin1996,EOGM},
or in the link field sector.
This is likely to allow unusual hierarchies of masses,
as in the (Extra)Ordinary Gauge Mediation models studied in \cite{EOGM}.
Finally, it would also be nice to consider richer link fields sectors,
e.g. larger quivers, of the type analyzed in~\cite{Csaki,McGarrie},
and to study in detail the regime where the messenger scale $M$
is not comparable to the massive gauge particles scale $m_v$.

\bigskip
\noindent{\bf Note Added:}
{}For sufficiently small $\tan\beta$,
our models may have the Tevatron signatures of promptly-decaying
slepton co-NLSPs, studied recently in~\cite{Ruderman:2010kj};
we thank Nathan Seiberg for pointing out this work to us.

\bigskip
\noindent{\bf Acknowledgements:}
We thank Zohar Komargodski and Vassilis Spanos for discussions.
This work was supported in part
by the BSF -- American-Israel Bi-National Science Foundation,
by a center of excellence supported by the Israel Science Foundation
(grant number 1665/10), DIP grant H.52, and the Einstein Center at the Hebrew University.

\begin{table*}[h]
\begin{center}
\begin{tabular}  {|l|l|l|l|} \hline
$N_{\rm mess}=1$ & $M=3.2 \times 10^5$ &  $M=10^8$ & $M=10^{15}$ \\
$ \tan \, \beta = 8$  & $\Lambda=0.99 M $ &  $\Lambda=2.7 \times 10^5$ & $\Lambda=2.0 \times 10^{5}$ \\
  $\alpha_{g_2}^{-1} =10  $   & $v=1.3 \times 10^5$ & $v=3.9 \times 10^7$ & $v=3.6 \times 10^{14}$ \\
 \hline
 $(y_1,y_2,y_3)$  & $(1, 1.10, 1.62)$ & $(1, 1.09, 1.38)$ & $(1,1.03,1.05)$ \\
  $(s(1),s(2),s(3))$  & $(0.02, 0.03, 0.08)$ & $(0.10,0.12,0.16)$ & $(0.10,0.11,0.11)$ \\
 \hline
$(M_1,M_2,M_3)$ &$(665, 1144, 2360)$  &$(477, 753, 1274)$ & $ (571, 625, 658)$  \\
$(m_Q,m_u,m_d)$ &$(847, 829, 827)$  &$(898, 847, 842)$ & $(439, 382, 364)$  \\
$(m_L,m_e)$ &$(186, 79)$  &$(328, 169)$ & $ (271, 202)$  \\
 $(\mu,B \mu)$ & $(687, 881^2)$ &$(821, 894^2)$ & $(787, 745^2)$ \\
\hline
$m_{\tilde{g}}$ &$2773$  &$1855$ & $1413$  \\
\hline
$m_{\tilde{\chi}_0}$ & $(571, 681, 683, 1115)$  &$(362, 672, 799, 829)$ & $(266, 502, 792, 804)$  \\
   \hline
$m_{\tilde{\chi}_{\pm}}$   & $(672, 1115)$ & $(672, 828)$ & $(503, 804)$  \\
\hline
$(m_{\tilde{u}_L}, m_{\tilde{d}_L})$ & $(1810, 1811)$ &$(1625, 1627)$  & $(1333, 1335)$  \\
$(m_{\tilde{u}_R},m_{\tilde{d}_R})$ & $(1775, 1775)$ &$(1573, 1571)$ & $(1270, 1262)$  \\
$(m_{\tilde{t}_1},m_{\tilde{t}_2})$ & $(1649, 1774)$ & $(1382, 1562)$ & $(1018, 1258)$ \\
$(m_{\tilde{b}_1},m_{\tilde{b}_2})$ & $(1753, 1774)$ &$(1539, 1569)$ & $(1224, 1259)$  \\
\hline
$(m_{\tilde{e}_R}, m_{\tilde{e}_L},m_{\tilde{\nu}_e})$ &
$(168, 400, 392)$ & $(219, 461, 454)$   & $(294, 484, 477)$  \\
$(m_{\tilde{\tau}_1},m_{\tilde{\tau}_2},m_{\tilde{\nu}_\tau})$ &
$(165, 400, 391)$ & $(216, 461, 453)$ & $(289, 484, 477)$ \\
\hline
  $m_{h_0}$ & $116$ &$116$  & $116$ \\
    $(m_{H_0},m_{A_0},m_{H_{\pm}})$ & $(783, 783, 787)$ &$(919, 919, 923)$ & $(924, 923, 927)$ \\
      \hline
 $(\mu,B \mu)$ & $(671, 815^2)$ &$(793,937^2)$ & $(787, 933^2)$ \\
   \hline
\end{tabular}
\caption{\footnotesize Sparticle masses in some numerical examples
with $\mu>0$ and $\tan \beta = 8$.
All the masses are in GeV.
The notation $s(r)$ is a short version for $s(x, y_r)$.
The input masses $(M_r,m_{Q,u,d,L,e})$ are at the messenger scale $M$;
$(\mu, B \mu)$ under these soft masses are
also evaluated at $M$.
 In the last line, the parameters $(\mu, B \mu)$ are evaluated
 at $Q = \sqrt{m_{\tilde{t}_1} m_{\tilde{t}_2}  }$. }
\label{masses-example-tan8}
\end{center}
\end{table*}

\begin{table*}[h]
\begin{center}
\begin{tabular}  {|l|l|l|l|} \hline
$N_{\rm mess}=1$ & $M=2.4 \times 10^5$ &  $M=10^8$ & $M=10^{15}$ \\
$ \tan \, \beta = 20$ & $\Lambda=0.99 M $ &  $\Lambda=2.1 \times 10^5$ & $\Lambda=1.5 \times 10^{5}$ \\
  $\alpha_{g_2}^{-1} =10  $ & $v=9.6 \times 10^4$ & $v=3.9 \times 10^7$ & $v=3.7 \times 10^{14}$ \\
 \hline
 $(y_1,y_2,y_3)$  & $(1,1.10,1.66)$ & $(1, 1.10, 1.39)$ & $(1,1.03,1.05)$ \\
   $(s(1),s(2),s(3))$   & $(0.02,0.03,0.09)$ & $(0.10, 0.12, 0.16)$ & $(0.10,0.11,0.11)$ \\
 \hline
$(M_1,M_2,M_3)$ &$(497,862,1808)$  &$(372, 589, 999)$ & $ (430,472,497)$  \\
$(m_Q,m_u,m_d)$ &$(665,652,651)$  &$(707, 668, 663)$ & $(332,289,275)$  \\
$(m_L,m_e)$ &$(140,59)$  &$(257, 131)$ & $ (205,152)$  \\
 $(\mu,B \mu)$ & $(534,736^2)$ &$(661, 598^2)$ & $(610,-176^2)$ \\
\hline
$m_{\tilde{g}}$ &$2131$  &$1475$ & $1089$  \\
\hline
$m_{\tilde{\chi}_0}$ & $(425, 526, 528, 845)$  &$(281, 520, 637, 666)$ & $(198, 375, 606, 618)$  \\
   \hline
$m_{\tilde{\chi}_{\pm}}$   & $(516, 845)$ & $(520, 665)$ & $(375, 618)$  \\
\hline
$(m_{\tilde{u}_L}, m_{\tilde{d}_L})$ & $(1405, 1407)$ &$(1298, 1300)$  & $(1027, 1030)$  \\
$(m_{\tilde{u}_R},m_{\tilde{d}_R})$ & $(1380, 1380)$ &$(1258, 1256)$ & $(981, 975)$  \\
$(m_{\tilde{t}_1},m_{\tilde{t}_2})$ & $(1280, 1383)$ & $(1102, 1251)$ & $(781, 978)$ \\
$(m_{\tilde{b}_1},m_{\tilde{b}_2})$ & $(1354, 1377)$ &$(1218, 1249)$ & $(929, 965)$  \\
\hline
$(m_{\tilde{e}_R}, m_{\tilde{e}_L},m_{\tilde{\nu}_e})$ &
$(129, 302, 292)$ & $(174, 363, 354)$   & $(223, 368, 359)$  \\
$(m_{\tilde{\tau}_1},m_{\tilde{\tau}_2},m_{\tilde{\nu}_\tau})$ &
$(106, 309, 291)$ & $(150, 368, 352)$ & $(197, 371, 355)$ \\
\hline
  $m_{h_0}$ & $116$ &$116$  & $116$ \\
    $(m_{H_0},m_{A_0},m_{H_{\pm}})$ & $(566, 566, 572)$ &$(689, 689, 694)$ & $(660, 660, 666)$ \\
      \hline
 $(\mu,B \mu)$ & $(519, 641^2)$ &$(631, 735^2)$ & $(600, 687^2)$ \\
   \hline
\end{tabular}
\caption{\footnotesize Sparticle masses in some numerical examples
with $\mu>0$ and $\tan \beta = 20$.
}
\label{masses-example-tan20}
\end{center}
\end{table*}

\begin{table*}[h]
\begin{center}
\begin{tabular}  {|l|l|l|l|} \hline
$N_{\rm mess}=1$ & $M=4.3 \times 10^5$ &  $M=10^8$ & $M=10^{15}$ \\
$ \tan \, \beta = 50$ & $\Lambda=0.99 M $ &  $\Lambda=3.2 \times 10^5$ & $\Lambda=2.2  \times 10^{5}$ \\
 $\alpha_{g_2}^{-1} =10  $    & $v=1.7 \times 10^5$ & $v=3.9 \times 10^7$ & $v=3.6   \times 10^{14}$ \\
 \hline
 $(y_1,y_2,y_3)$  & $(1, 1.10, 1.58)$ & $(1, 1.09, 1.38)$ & $(1,1.03,1.04)$ \\
   $(s(1),s(2),s(3))$    & $(0.02,0.03,0.08)$ & $(0.10, 0.12, 0.16)$ & $(0.10,0.11,0.11)$ \\
 \hline
$(M_1,M_2,M_3)$ &$(897, 1529, 3101)$  &$(564, 887, 1499)$ & $ (627, 684, 721)$  \\
$(m_Q,m_u,m_d)$ &$(1086, 1061, 1059)$  &$(1055, 995, 988)$ & $(481, 419, 399)$  \\
$(m_L,m_e)$ &$(248, 107)$  &$(387, 199)$ & $ (297, 221)$  \\
 $(\mu,B \mu)$ & $(896, 856^2)$ &$(1004, -1631^2)$ & $(982, -2811^2)$ \\
\hline
$m_{\tilde{g}}$ &$3634$  &$2163$ & $1539$  \\
\hline
$m_{\tilde{\chi}_0}$ & $(771, 844, 853, 1485)$  &$(432, 797, 894, 927)$ & $(294, 556, 831, 841)$  \\
   \hline
$m_{\tilde{\chi}_{\pm}}$   & $(839, 1485)$ & $(797, 926)$ & $(556, 841)$  \\
\hline
$(m_{\tilde{u}_L}, m_{\tilde{d}_L})$ & $(2347, 2348)$ &$(1892, 1893)$  & $(1453, 1455)$  \\
$(m_{\tilde{u}_R},m_{\tilde{d}_R})$ & $(2299, 2298)$ &$(1830, 1826)$ & $(1384, 1374)$  \\
$(m_{\tilde{t}_1},m_{\tilde{t}_2})$ & $(2143, 2252)$ & $(1614, 1762)$ & $(1117, 1310)$ \\
$(m_{\tilde{b}_1},m_{\tilde{b}_2})$ & $(2195, 2247)$ &$(1694, 1758)$ & $(1219, 1298)$  \\
\hline
$(m_{\tilde{e}_R}, m_{\tilde{e}_L},m_{\tilde{\nu}_e})$ &
$(223, 532, 526)$ & $(258, 541, 535)$   & $(322, 529, 523)$  \\
$(m_{\tilde{\tau}_1},m_{\tilde{\tau}_2},m_{\tilde{\nu}_\tau})$ &
$(98, 548, 518)$ & $(99, 552, 520)$ & $(100, 522, 490)$ \\
\hline
  $m_{h_0}$ & $117$ &$117$  & $117$ \\
    $(m_{H_0},m_{A_0},m_{H_{\pm}})$ & $(601, 602, 608)$ &$(672, 673, 677)$ & $(625, 625, 630)$ \\
      \hline
 $(\mu,B \mu)$ & $(836, 1008^2)$ &$(891, 909^2)$ & $(829, 758^2)$ \\
   \hline
\end{tabular}
\caption{\footnotesize Sparticle masses in some numerical examples
with $\mu>0$ and $\tan \beta = 50$.}
\label{masses-example-tan50}
\end{center}
\end{table*}

\begin{table*}[h]
\begin{center}
\begin{tabular}  {|l|l|l|l|} \hline
$N_{\rm mess}=5$ & $M=7.0 \times 10^4$ &  $M=10^8$ & $M=10^{15}$ \\
$ \tan \, \beta = 20$ & $\Lambda=0.99 M $ &  $\Lambda=5 \times 10^4$ & $\Lambda=3.4 \times 10^{4}$ \\
  $\alpha_{g_2}^{-1} =10  $     & $v=2.8 \times 10^4$ & $v=3.9 \times 10^7$ & $v=3.6 \times 10^{14}$ \\
 \hline
 $(y_1,y_2,y_3)$  & $(1,1.10,1.73)$ & $(1,1.10,1.39)$ & $(1,1.03,1.05)$ \\
  $(s(1),s(2),s(3))$    & $(0.02,0.03,0.10)$ & $(0.10,0.12,0.16)$ & $(0.10,0.11,0.11)$ \\
 \hline
$(M_1,M_2,M_3)$ &$(706,1240,2709)$  &$(443, 700, 1185)$ & $ (487, 534, 562)$  \\
$(m_Q,m_u,m_d)$ &$(462, 454, 453)$  &$(374, 353, 351)$ & $(168, 146, 139)$  \\
$(m_L,m_e)$ &$(90, 38)$  &$(137, 70)$ & $ (104, 77)$  \\
 $(\mu,B \mu)$ & $(496, 812^2)$ &$(647, 576^2)$ & $(655, -204^2)$ \\
\hline
$m_{\tilde{g}}$ &$3005$  &$1705$ & $1212$  \\
\hline
$m_{\tilde{\chi}_0}$ & $(481, 495, 645, 1212)$  &$(334, 578, 626, 696)$ & $(225, 424, 650, 664)$  \\
   \hline
$m_{\tilde{\chi}_{\pm}}$   & $(490, 1212)$ & $(578, 695)$ & $(424, 664)$  \\
\hline
$(m_{\tilde{u}_L}, m_{\tilde{d}_L})$ & $(1577, 1579)$ &$(1325, 1327)$  & $(1102, 1105)$  \\
$(m_{\tilde{u}_R},m_{\tilde{d}_R})$ & $(1548, 1548)$ &$(1294, 1293)$ & $(1060, 1057)$  \\
$(m_{\tilde{t}_1},m_{\tilde{t}_2})$ & $(1459, 1558)$ & $(1145, 1291)$ & $(855, 1052)$ \\
$(m_{\tilde{b}_1},m_{\tilde{b}_2})$ & $(1530, 1546)$ &$(1253, 1285)$ & $(1004, 1045)$  \\
\hline
$(m_{\tilde{e}_R}, m_{\tilde{e}_L},m_{\tilde{\nu}_e})$ &
$(144, 345, 335)$ & $(149, 333, 323)$   & $(200, 360, 351)$  \\
$(m_{\tilde{\tau}_1},m_{\tilde{\tau}_2},m_{\tilde{\nu}_\tau})$ &
$(129, 348, 334)$ & $(122, 339, 321)$ & $(171, 364, 347)$ \\
\hline
  $m_{h_0}$ & $116$ &$116$  & $116$ \\
     $(m_{H_0},m_{A_0},m_{H_{\pm}})$ & $(560, 560, 566)$ &$(663, 663, 668)$ & $(693, 693, 698)$ \\
      \hline
 $(\mu,B \mu)$ & $(485,652^2)$ &$(618, 717^2)$ & $(645, 723^2)$ \\
   \hline
\end{tabular}
\caption{\footnotesize Sparticle masses in some numerical examples
with $\mu>0$, $\tan \beta = 20$ and $N_{\rm mess}=5$.}
\label{masses-example-nmess5}
\end{center}
\end{table*}

\begin{table*}[h]
\begin{center}
\begin{tabular}  {|l|l|l|l|} \hline
$N_{\rm mess}=1$ & $M=1.5 \times 10^5$ &  $M=10^8$ & $M=10^{15}$ \\
$ \tan \, \beta = 20$  & $\Lambda=0.99 M $ &  $\Lambda=1.6 \times 10^5$ & $\Lambda=1.5 \times 10^{5}$ \\
 \hline
$(M_1,M_2,M_3)$ &$(307, 538, 1161)$  &$(282, 448, 764)$ & $ (427, 469, 495)$  \\
$(m_Q,m_u,m_d)$ &$(1336, 1270, 1264)$  &$(1363, 1264, 1251)$ & $(996, 867, 824)$  \\
$(m_L,m_e)$ &$(453, 224)$  &$(570, 309)$ & $ (621, 468)$  \\
 $(\mu,B \mu)$ & $(542, 785^2)$ &$(820, 874^2)$ & $(797, 585^2)$ \\
\hline
$m_{\tilde{g}}$ &$1469$  &$1214$ & $1125$  \\
\hline
$m_{\tilde{\chi}_0}$ & $(270, 478, 537, 593)$  &$(216, 418, 791, 799)$ & $(200, 385, 789, 796)$  \\
   \hline
$m_{\tilde{\chi}_{\pm}}$   & $(477, 591)$ & $(418, 799)$ & $(385, 796)$  \\
\hline
$(m_{\tilde{u}_L}, m_{\tilde{d}_L})$ & $(1560, 1562)$ &$(1593, 1595)$  & $(1372, 1374)$  \\
$(m_{\tilde{u}_R},m_{\tilde{d}_R})$ & $(1498, 1494)$ &$(1500, 1490)$ & $(1260, 1229)$  \\
$(m_{\tilde{t}_1},m_{\tilde{t}_2})$ & $(1384, 1513)$ & $(1278, 1493)$ & $(948, 1246)$ \\
$(m_{\tilde{b}_1},m_{\tilde{b}_2})$ & $(1479, 1504)$ &$(1462, 1490)$ & $(1194, 1234)$  \\
\hline
$(m_{\tilde{e}_R}, m_{\tilde{e}_L},m_{\tilde{\nu}_e})$ &
$(240, 483, 476)$ & $(323, 600, 594)$   & $(494,686, 681)$  \\
$(m_{\tilde{\tau}_1},m_{\tilde{\tau}_2},m_{\tilde{\nu}_\tau})$ &
$(229, 483, 474)$ & $(304, 599, 590)$ & $(464, 679, 672)$ \\
\hline
  $m_{h_0}$ & $116$ &$116$  & $116$ \\
    $(m_{H_0},m_{A_0},m_{H_{\pm}})$ & $(677, 677, 683)$ &$(939, 939, 942)$ & $(984, 985, 988)$ \\
      \hline
 $(\mu,B \mu)$ & $(528, 733^2)$ &$(783, 977^2)$ & $(784, 1011^2)$ \\
   \hline
\end{tabular}
\caption{\footnotesize Sparticle masses in some numerical examples in Minimal Gauge Mediation,
with $\mu>0$ and $\tan \beta = 20$.
}
\label{masses-example-mgm}
\end{center}
\end{table*}

\begin{table*}[h]
\begin{center}
\begin{tabular}  {|l|l|l|l|} \hline
$N_{\rm mess}=5$ & $M=5.0 \times 10^4$ &  $M=10^8$ & $M=10^{15}$ \\
$ \tan \, \beta = 20$  & $\Lambda=0.99 M $ &  $\Lambda=4 \times 10^4$ & $\Lambda=3.4 \times 10^{4}$ \\
 \hline
$(M_1,M_2,M_3)$ &$(501, 888, 1981)$  &$(354, 561, 953)$ & $ (486, 533, 561)$  \\
$(m_Q,m_u,m_d)$ &$(1016, 968, 964)$  &$(761, 706, 698)$ & $(505, 439, 417)$  \\
$(m_L,m_e)$ &$(334, 163)$  &$(319, 173)$ & $ (315, 238)$  \\
 $(\mu,B \mu)$ & $(441, 738^2)$ &$(643, 605^2)$ & $(704, 224^2)$ \\
\hline
$m_{\tilde{g}}$ &$2218$  &$1418$ & $1226$  \\
\hline
$m_{\tilde{\chi}_0}$ & $(408, 441, 478, 880)$  &$(267, 497, 621, 647)$ & $(226, 428, 698, 709)$  \\
   \hline
$m_{\tilde{\chi}_{\pm}}$   & $(433, 880)$ & $(497, 646)$ & $(429, 709)$  \\
\hline
$(m_{\tilde{u}_L}, m_{\tilde{d}_L})$ & $(1515, 1517)$ &$(1290, 1292)$  & $(1196, 1199)$  \\
$(m_{\tilde{u}_R},m_{\tilde{d}_R})$ & $(1469, 1466)$ &$(1241, 1237)$ & $(1134, 1123)$  \\
$(m_{\tilde{t}_1},m_{\tilde{t}_2})$ & $(1383, 1487)$ & $(1082, 1239)$ & $(901, 1121)$ \\
$(m_{\tilde{b}_1},m_{\tilde{b}_2})$ & $(1454, 1476)$ &$(1206, 1232)$ & $(1078, 1112)$  \\
\hline
$(m_{\tilde{e}_R}, m_{\tilde{e}_L},m_{\tilde{\nu}_e})$ &
$(195, 411, 403)$ & $(205, 402, 394)$   & $(300, 465, 458)$  \\
$(m_{\tilde{\tau}_1},m_{\tilde{\tau}_2},m_{\tilde{\nu}_\tau})$ &
$(186, 412, 402)$ & $(186, 405, 392)$ & $(274, 464, 452)$ \\
\hline
  $m_{h_0}$ & $116$ &$116$  & $116$ \\
    $(m_{H_0},m_{A_0},m_{H_{\pm}})$ & $(561, 561, 567)$ &$(695, 695, 699)$ & $(786, 785, 790)$ \\
      \hline
 $(\mu,B \mu)$ & $(432, 636^2)$ &$(614, 739^2)$ & $(693, 815^2)$ \\
   \hline
\end{tabular}
\caption{\footnotesize Sparticle masses in some numerical examples in Minimal Gauge Mediation,
with $\mu>0$, $\tan \beta = 20$ and $N_{\rm mess}=5$.}
\label{masses-example-mgm-nmess5}
\end{center}
\end{table*}


\begin{thebibliography}{99}

\bibitem{Cheng}
  H.~C.~Cheng, D.~E.~Kaplan, M.~Schmaltz and W.~Skiba,
  Phys.\ Lett.\  B {\bf 515} (2001) 395
  [arXiv:hep-ph/0106098].

\bibitem{Csaki}
  C.~Csaki, J.~Erlich, C.~Grojean and G.~D.~Kribs,
  Phys.\ Rev.\  D {\bf 65} (2002) 015003
  [arXiv:hep-ph/0106044].

\bibitem{KKS}
  D.~E.~Kaplan, G.~D.~Kribs and M.~Schmaltz,
  Phys.\ Rev.\  D {\bf 62}, 035010 (2000)
  [arXiv:hep-ph/9911293].

\bibitem{Chacko}
  Z.~Chacko, M.~A.~Luty, A.~E.~Nelson and E.~Ponton,
  JHEP {\bf 0001} (2000) 003
  [arXiv:hep-ph/9911323].

\bibitem{GGM}
  P.~Meade, N.~Seiberg and D.~Shih,
  Prog.\ Theor.\ Phys.\ Suppl.\  {\bf 177} (2009) 143
  [arXiv:0801.3278 [hep-ph]].

 \bibitem{dsfs1}
  A.~De Simone, J.~Fan, M.~Schmaltz and W.~Skiba,
  Phys.\ Rev.\  D {\bf 78} (2008) 095010
  [arXiv:0808.2052 [hep-ph]].

\bibitem{dsfs2}
  A.~De Simone, J.~Fan, V.~Sanz and W.~Skiba,
  Phys.\ Rev.\  D {\bf 80} (2009) 035010
  [arXiv:0903.5305 [hep-ph]].

\bibitem{gkk}
  D.~Green, A.~Katz and Z.~Komargodski,
  arXiv:1008.2215 [hep-th].

   \bibitem{mggmsoft}
   R.~Auzzi and A.~Giveon,
  JHEP {\bf 1010} (2010) 088
  [arXiv:1009.1714 [hep-ph]].

\bibitem{McGarrie}
   M.~McGarrie and R.~Russo,
  Phys.\ Rev.\  D {\bf 82}, 035001 (2010)
  [arXiv:1004.3305 [hep-ph]];
   M.~McGarrie,
  arXiv:1009.0012 [hep-ph].

\bibitem{Sudano}
 M.~Sudano,
  arXiv:1009.2086 [hep-ph].

\bibitem{Martin1996}
  S.~P.~Martin,
  Phys.\ Rev.\  D {\bf 55} (1997) 3177
  [arXiv:hep-ph/9608224].

\bibitem{Dimopoulos1996}
  S.~Dimopoulos, G.~F.~Giudice and A.~Pomarol,
  Phys.\ Lett.\  B {\bf 389}, 37 (1996)
  [arXiv:hep-ph/9607225].

\bibitem{veltman}
  J.~van der Bij and M.~J.~G.~Veltman,
  Nucl.\ Phys.\  B {\bf 231} (1984) 205.

\bibitem{softsusy}
  B.~C.~Allanach,
  Comput.\ Phys.\ Commun.\  {\bf 143} (2002) 305
  [arXiv:hep-ph/0104145].

\bibitem{martin-review}
S.~P.~Martin,
  arXiv:hep-ph/9709356.

 \bibitem{mv}
  S.~P.~Martin and M.~T.~Vaughn,
  Phys.\ Rev.\  D {\bf 50} (1994) 2282
  [Erratum-ibid.\  D {\bf 78} (2008) 039903]
  [arXiv:hep-ph/9311340].

 \bibitem{pbmz}
  D.~M.~Pierce, J.~A.~Bagger, K.~T.~Matchev and R.~j.~Zhang,
  Nucl.\ Phys.\  B {\bf 491} (1997) 3
  [arXiv:hep-ph/9606211].

\bibitem{mgma}
  J.~A.~Bagger, K.~T.~Matchev, D.~M.~Pierce and R.~j.~Zhang,
  Phys.\ Rev.\  D {\bf 55} (1997) 3188
  [arXiv:hep-ph/9609444].

\bibitem{mgmb}
  S.~Dimopoulos, S.~D.~Thomas and J.~D.~Wells,
  Nucl.\ Phys.\  B {\bf 488} (1997) 39
  [arXiv:hep-ph/9609434].

\bibitem{ADJK}
  S.~Abel, M.~J.~Dolan, J.~Jaeckel, V.~V.~Khoze, 
  arXiv:1009.1164 [hep-ph].


\bibitem{GMGM}
 D.~Marques,
  JHEP {\bf 0903} (2009) 038
  [arXiv:0901.1326 [hep-ph]];
  T.~T.~Dumitrescu, Z.~Komargodski, N.~Seiberg and D.~Shih,
  JHEP {\bf 1005} (2010) 096
  [arXiv:1003.2661 [hep-ph]].

\bibitem{EOGM}
C.~Cheung, A.~L.~Fitzpatrick and D.~Shih,
 JHEP {\bf 0807}, 054 (2008)
 [arXiv:0710.3585 [hep-ph]].

\bibitem{Ruderman:2010kj}
  J.~T.~Ruderman and D.~Shih,
  arXiv:1009.1665 [hep-ph].



\end{thebibliography}
\end{document}